\documentclass[prl,twocolumn,english,superscriptaddress,floatfix]{revtex4}
\usepackage{graphicx}
\usepackage{bm, amsmath, amssymb}
 \usepackage{soul}
\usepackage{pdfpages}
\usepackage[T1]{fontenc}
\usepackage[latin9]{inputenc}
\usepackage{amstext}
\usepackage{bbold}
\usepackage{esint}
\newcommand{\nbar}[0]{\bar{n}}

\newcommand{\op}[1]{\hat{ #1}}                
\def\be{\begin{equation}}
\def\ee{\end{equation}}
\def\bea{\begin{eqnarray}}
\def\eea{\end{eqnarray}}

\usepackage{babel}

\begin{document}

\title{Prediction and retrodiction for a continuously monitored superconducting qubit}

\author{D. Tan}
\affiliation{Department of Physics, Washington University, St.\ Louis, Missouri 63130}
\author{S. J. Weber}
\affiliation{Quantum Nanoelectronics Laboratory, Department of Physics, University of California, Berkeley CA 94720}
\author{I. Siddiqi}
\affiliation{Quantum Nanoelectronics Laboratory, Department of Physics, University of California, Berkeley CA 94720}
\author{K. M\o lmer}
\affiliation{Department of Physics and Astronomy, Aarhus University, Ny Munkegade 120, DK-8000 Aarhus C, Denmark}
\author{K. W. Murch*}
\affiliation{Department of Physics, Washington University, St.\ Louis, Missouri 63130}

\date{\today}

\begin{abstract}
The quantum state of a superconducting transmon qubit inside a three-dimensional cavity is monitored by transmission of a microwave field through the cavity. 
The information inferred from the measurement record is incorporated in a density matrix $\rho_t$, which is conditioned on probe results until $t$, and in an auxiliary matrix $E_t$, which is conditioned on probe results obtained after $t$.  Here, we obtain these matrices from experimental data and we illustrate their application to predict and retrodict the outcome of weak and strong qubit measurements.
\end{abstract}

\maketitle

In quantum mechanics, predictions about the outcome of experiments are given by Born's rule which for a state vector $|\psi_i\rangle$ provides the probability $P(a)=|\langle a|\psi_i\rangle|^2$ that a measurement of an observable $\hat{A}$ with eigenstates $|a\rangle$ yields one of the eigenvalues $a$. As a consequence of the measurement, the quantum state is projected into the state $|a\rangle$.  Yet, after this measurement, further probing of the system is possible, and the probability that the quantum system yields outcome $a$  and is subsequently detected in a final state $|\psi_f\rangle$ factors into the product $|\langle\psi_f| a\rangle|^2 |\langle a|\psi_i\rangle|^2$. Considering initial and final states raises the issue of post-selection in quantum measurements: What is the probability that the result of the measurement of $\hat{A}$ {\it was} $a$, if we consider only the selected measurement events where the initial state was $|\psi_i\rangle$ and the final state was $|\psi_f\rangle$?
The answer is known as the Aharonov-Bergmann-Lebowitz rule \cite{ahar64},
\begin{equation}
P_{ABL}(a)= \frac{P(f,a|i)}{\sum_{a'}P(f,a'|i)} = \frac{|\langle\psi_f| a\rangle\langle a|\psi_i\rangle|^2}{\sum_{a'} |\langle\psi_f| a'\rangle\langle a'|\psi_i\rangle|^2} \label{eq:abl}
\end{equation}
and it differs from Born's rule, which takes into account only knowledge about the state prior to the measurement.

While it is natural that full measurement records reveal more information about the state of a physical system at a given time $t$ than data obtained only until that time, the interpretation of the time symmetric influences from the future and from the past measurement events on $P_{ABL}$ has stimulated some debate, see for example \cite{wana55,ahar64,ahar10, ahar11, ahar11_2, vaid13}. Meanwhile, probabilistic state assignments and correlations observed in atomic, optical and solid state experiments have been conveniently understood in relation to post-selection \cite{meschede,gogg11,wald11,groe13,camp13}, and precision probing theories \cite{mank09,mank09pra,mank11,arme09,whea10,ryba14} have incorporated full measurement records.

In this letter, we consider a superconducting qubit that is subject to continuous monitoring and driven unitary evolution. We make use of the full measurement record and examine how measurements before time $t$ can be used to make predictions, while measurements after time $t$ can be used to make retrodictions about measurements at time $t$. We then consider a recent generalization \cite{gamm13} of Eq.(\ref{eq:abl}) to the case of continuously monitored and evolving mixed states. Our experiments  verify the predictions of both projective and weak (weak value)  measurements conditioned on full measurement records.  These predictions are more confident and nontrivially different from predictions based only on the measurement record up to time $t$.    

To analyze non-pure states and partial measurements, we represent our system by a density matrix $\rho$, and measurements  by the theory of positive operator-valued measures (POVM) which yields the probability $P(m)=\mathrm{Tr}(\Omega_m \rho \Omega_m^\dagger)$ for outcome $m$, and the associated back action on the quantum state, $\rho \rightarrow \Omega_m \rho \Omega_m^\dagger/P(m)$, where the operators $\Omega_m$ obey  $\sum_m \Omega^\dagger_m \Omega_m = \hat{I}$. When $\Omega_a = |a\rangle \langle a|$ is a projection operator and $\rho = |\psi\rangle\langle \psi|$, the theory of POVMs is in agreement with Born's rule.

For systems subject to unitary and dissipative time evolution along with continuous monitoring before and after a measurement described by operator $\Omega_m$, one can show \cite{gamm13} that,
\begin{eqnarray}
P_p(m)  = \frac{\mathrm{Tr}(\Omega_m \rho_t \Omega_m^\dagger E_t)}{\sum_m \mathrm{Tr}(\Omega_m \rho_t \Omega_m^\dagger E_t)}, \label{eq:pqs}
\end{eqnarray}
where $\rho_t$ is the system density matrix at time $t$, conditioned on previous measurement outcomes, and propagated forward in time until time $t$, while $E_t$ is a matrix which is propagated backwards in time in a similar manner and accounts for the time evolution and measurements obtained after time $t$. The subscript $\cdot_p$ denotes ``past'', and in \cite{gamm13} it was proposed that, if $t$ is in the past, the pair of matrices $(\rho_t,E_t)$, rather than only $\rho_t$, is the appropriate object to associate with the state of a quantum system at time $t$. We observe that for the case of pure states and projective measurements, $P_p(m)$ in (\ref{eq:pqs}) acquires the form of Eq.(1) with $\rho_t = |\psi_i\rangle\langle \psi_i|$ and $E_t=|\psi_f\rangle\langle \psi_f|$.

Here, we make use of the full measurement record to compute the matrices $\rho_t$ and $E_t$,  and analyze how they, through application of  Eq.(\ref{eq:pqs}) yield more confident predictions for measurements on the system. For imperfect measurement efficiency, non-pure states, and measurements that do not commute with the system evolution, the predictions of Eq.(\ref{eq:pqs})  vary dramatically from those based on $\rho$ alone \cite{murc13traj,webe14}.


 \begin{figure}\begin{center}
\includegraphics[angle = 0, width =.48\textwidth]{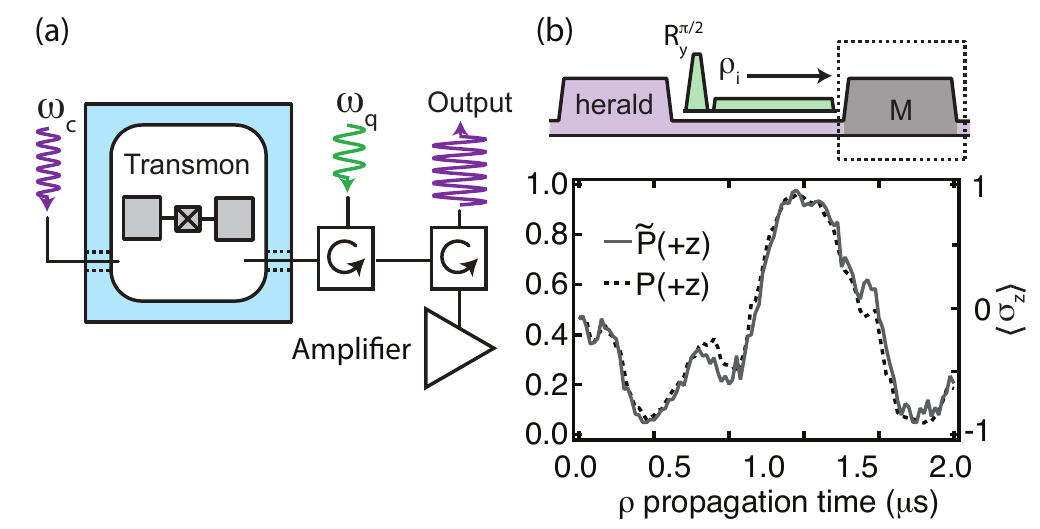}
\end{center}
\caption{ \label{fig1} Time evolution in a monitored system.  
(a) Simplified experimental setup consisting of a transmon circuit coupled to a waveguide cavity.  (b) We prepare the qubit in an initial state ($\mathrm{Tr}(\rho_i \sigma_x) \simeq +1$) and propagate $\rho$ forward in time, which makes accurate predictions about a final projective measurement (in the $\sigma_z$ basis) labeled $M$. The dashed line is the prediction based on a single quantum trajectory, and the solid line is the result from projective measurements on an ensemble of experiments that have similar values of $\rho_t$.}
\end{figure}

Our experiment, illustrated in figure 1a, is composed of a superconducting transmon circuit dispersively coupled to a wave-guide cavity \cite{koch07,paik113D}.  The two lowest energy levels of the transmon form a qubit with transition frequency $\omega_q/2 \pi = 4.0033$ GHz.  The dispersive coupling between the transmon qubit and the cavity is given by an interaction Hamiltonian, $H_\mathrm{int.}  = -\hbar \chi a^\dagger a \sigma_z$, where $\hbar$ is the reduced Plank's constant, $a^\dagger (a)$ is the creation (annihilation) operator for the cavity mode at frequency $\omega_c/2\pi =6.9914$ GHz, $\chi/2\pi =-0.425$ MHz is the dispersive coupling rate, and $\sigma_z$ is the qubit Pauli operator that acts on the qubit in the energy basis.  A microwave tone that probes the cavity with an average intracavity photon number $\nbar = \langle a^\dagger a\rangle$ thus acquires a qubit-state-dependent phase shift. Since $2|\chi|\ll\kappa$, where $\kappa/2\pi = 9.88$ MHz is the cavity linewidth,  qubit state information is encoded in one quadrature of the microwave signal.  We amplify this quadrature  of the signal with a near-quantum-limited Josephson parametric amplifier \cite{hatr11para}. After further amplification, the measurement signal is demodulated and digitized.  This setup allows variable strength measurements of the qubit state characterized by a measurement timescale $\tau$;  by binning the measurement signal in time steps $\delta t\ll \tau$ we execute weak measurements of the qubit state \cite{hatr13,murc13traj} while by integrating the measurement signal for a time $T\gg \tau$ we effectively accumulate  weak measurements in a  projective measurement \cite{vija11} of the qubit in the $\sigma_z$ basis.

Our experimental sequences begin with a projective measurement of the qubit in the $\sigma_z$ basis followed by a variable rotation of the qubit state  to prepare the qubit in an arbitrarily specified initial (nearly) pure state. 
Following this preparation, the qubit is subject to continuous rotations given by $H_\mathrm{R} = \hbar \Omega_R \sigma_y/2$, where $\Omega_R/2\pi =0.7$ MHz is the Rabi frequency, and continuous probing given by the measurement operator $\sqrt{k} \sigma_z$, where $k = 4\chi^2 \nbar/\kappa = 1/4\eta\tau$ parametrizes the measurement strength ($k/2\pi = 95$ kHz) and $\eta = 0.35$ is the quantum measurement efficiency \cite{supp}. During probing, we digitize the measurement signal $V_t$ in time steps $\delta t = 20$ ns.

The density matrix associated with a given measurement signal $V_t$ is obtained  by solving the stochastic master equation \cite{wisebook}:
\begin{multline}
\frac{d \rho}{dt} = -\frac{\mathrm{{\bf i}}}{\hbar}[H_\mathrm{R},\rho] + k(\sigma_z \rho \sigma_z - \rho) \\
+ 2 \eta k( \sigma_z \rho +\rho \sigma_z - 2 \mathrm{Tr}(\sigma_z \rho) \rho) V_t. \label{eq:rho}
\end{multline}
Here, the first two terms are the standard master equation in Lindblad form, and the last stochastic term updates the state based on the measurement result and leads to quantum trajectory solutions that are different for every repetition of the experiment. 


Let us first recall how the density matrix makes predictions about the outcome of measurements.
In figure 1b, we consider the probabilities $P(\pm z)$ for the outcome of the projective measurement operators $\Omega_{\pm z} = (\sigma_z \pm 1)/2$.
We prepare the initial state, $\mathrm{Tr}(\rho_i \sigma_x) \simeq +1$, by heralding the ground state and applying a $\pi/2$ rotation about the $y$ axis.  We then propagate $\rho_t$ forward from this initial state, and at each point in time we display the calculated $P(+z)=\mathrm{Tr}(\Omega_{+z}\rho_t \Omega^\dagger_{+z})$ \cite{supp}. By performing projective measurements of $\Omega_{\pm z} $ at time $t$ on an ensemble of experiments that have similar values of $\rho_t$ (within $\pm 0.02$) we obtain the corresponding experimental result $\tilde{P}(+z)$. We perform this analysis at different times and we observe close agreement between the single quantum trajectory prediction $P(+z)$ and the observed $\tilde{P}(+z)$. Note that the same procedure was  used to tomographically reconstruct and verify the quantum trajectory associated with the mean value  $\langle \sigma_z\rangle = 2 P(+z)-1$ in \cite{murc13traj,webe14}.


We now turn to the application of measurement data to retrodict the outcome of an already performed measurement. Eq.(\ref{eq:pqs}) applies for any set of POVM measurement operators $\Omega_m$ at time $t$, and accumulates the information retrieved from the later probing in the matrix $E_t$ that is propagated \emph{backwards} in time according to \cite{gamm13},
 \begin{multline}
\frac{d E}{dt} = \frac{\mathrm{{\bf i}}}{\hbar}[H_\mathrm{R},E]  +  k(\sigma_z E \sigma_z - E) \\
+2 \eta k( \sigma_z E +E \sigma_z - \mathrm{Tr}(\sigma_z E) E)V_{t-dt}. \label{eq:E}
\end{multline}

We assume that no measurements take place beyond the time $T$, leading to the final condition  $E_T=\hat{I}$  \cite{gamm13} (Note that $\mathrm{Tr}(E) = 2)$. If no measurements take place at all before $T$, for example because $\eta=0$, Eq.(\ref{eq:E}) yields a solution for $E(t)$ that remains proportional to the identity operator for all times, and Eq.(\ref{eq:pqs}) leads to the conventional expression that depends only on $\rho_t$.  

\begin{figure}\begin{center}
\includegraphics[angle = 0, width =.48\textwidth]{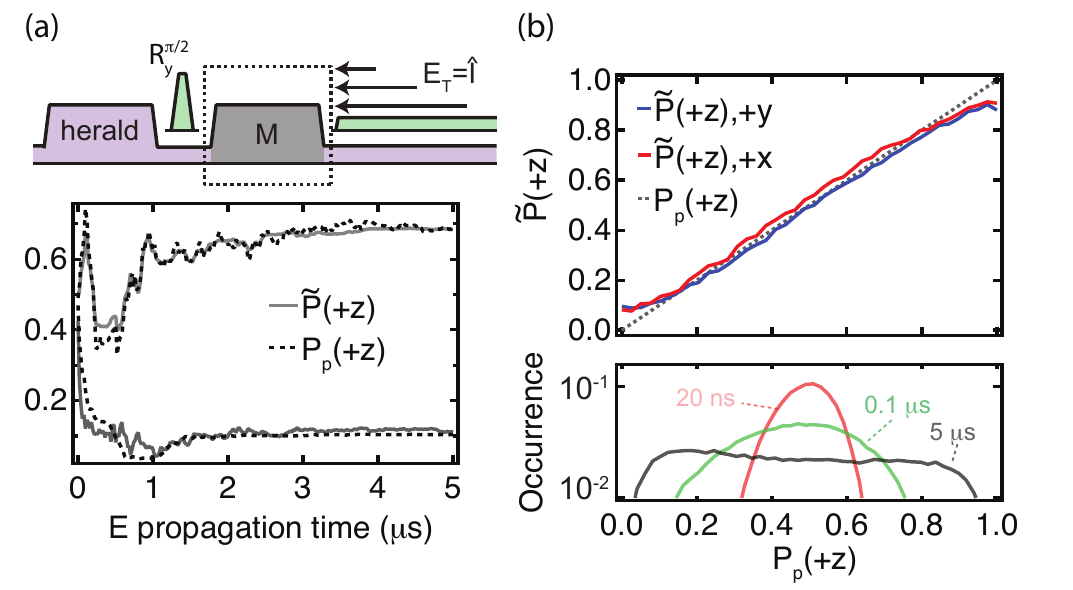}
\end{center}
\caption{ \label{fig2} Retrodiction in a monitored system.  
(a) To test retrodictions made by $E$ we prepare different states $\rho_i$ and conduct a subsequent projective  measurement $M$. We propagate $E$ backwards from the final state $E_T =\hat{I}$ to $E_0$ for variable periods of time ($T$). This yields a retrodiction (shown as dashed lines for two different experiments) for the outcome of $M$.  The solid line, which is based on an ensemble of experiments that yielded similar values of $E_0$ confirms the retrodictions based on the single measurement record. (b) We prepare two different initial states ($\mathrm{Tr}(\rho_i \sigma_x) \simeq +1$, red, $\mathrm{Tr}(\rho_i \sigma_y) \simeq +1$, blue), and compare the  retrodictions, $P_p(+z)$, based on $5\ \mu$s of probing, to the outcomes of measurements $M$ that yielded similar values of $E_0$.  In the lower panel, we display histograms of $P_p(+z)$ for different propagation times. As more of the  record is included, the retrodictions become more confident, taking values that are more often near 0 or 1. }
\end{figure}

In figure 2 we test the retrodictions made by $E$ and Eq.(\ref{eq:pqs}).  We examine different initial states, $\mathrm{Tr}(\rho_i \sigma_x) \simeq +1$, $\mathrm{Tr}(\rho_i \sigma_y) \simeq +1$, which are prepared by heralding the ground state and applying $\pi/2$ rotations about the $y$ and $-x$ axes respectively.  We propagate $E$ backwards from $E_T$ to $E_0$ to make a retrodiction about a projective measurement $M$.  Note that the initial states $\rho_i$ make ambiguous predictions about the outcome of $M$, $P(+z) = 1/2$, yet information is available after $M$ and by propagating $E$ for longer times, the retrodiction for the outcome of $M$ becomes more confident. 
  
  We verify that the retrodictions are correct by averaging the outcomes of many measurements $M$ that corresponded to similar values of $E_0$ to obtain an experimentally derived probability, $\tilde{P}(+z)$.    Figure 2a displays two sample trajectories for the retrodiction $P_p(+z)$ along with $\tilde{P}(+z)$. As more information is included, the retrodictions converge to fixed values. 
   Figure 2b displays the results of $3\times10^{5}$ experimental tests for the two different initial states $\rho_i$. For both initial states and for a wide range of measurement outcomes we are able to tomographically verify the retrodictions. We also display histograms of the different values $P_p(+z)$ for different propagation times of $E$. These show that as information is included in the propagation of $E$ the retrodictions become more confident.


Having verified the predictions based on $\rho$, and the retrodictions based on $E$, we now aim to illustrate the application of  $\rho$ and $E$ to create a \emph{smoothed} prediction (which uses both past and future information) for the outcome of a POVM measurement.  The POVM measurement that we consider is simply a short segment of the measurement signal received between $t$ and $t+\Delta t$ and is given by the measurement operators   \cite{wisebook, jaco06}, 
\begin{eqnarray}
\Omega_V = \left(2 \pi a^2 \right)^{-1/4} e^{(-(V- \sigma_z)^2/4a^2)}
\label{povm}
\end{eqnarray}
where,  $1/4a^2 =  k \eta \Delta t$.  The operators $\Omega_V$ satisfy $\int \Omega_V^\dagger \Omega_V dV = \op{I}$ as expected for POVMs, and if we assume that $\rho_t$ can be treated as a constant during $\Delta t$, the probability of the measurement yielding a value $V$ is $P(V) = \mathrm{Tr}(\Omega_V \rho_t \Omega_V^\dagger)$, which is the sum of two Gaussian distributions with variance $a^2$ centered at $+1$ and $-1$ and weighted by the populations $\rho_{00}$ and $\rho_{11}$ of the two qubit states. The $\sigma_z$ term in $\Omega_V$ causes the back action on the qubit degree of freedom, $\rho \rightarrow \Omega_V \rho \Omega_V^\dagger$, due to the readout of the measurement result $V$. If the effects of damping and the Rabi drive can be ignored during $\Delta t$, the operators (\ref{povm}) also describe a stronger measurement, yielding ultimately the limit where the two Gaussian distributions are disjoint, and the readout causes  projective back action of the qubit on one of its $\sigma_z$ eigenstates, with probabilities $\rho_{00}$ and $\rho_{11}$.

Since the system is also subject to probing and evolution after $t$, we now examine what smoothed predictions can be made for the outcome of the measurement $\Omega_V$  based on both earlier \emph{and} later probing. We must hence evaluate the conditioned density matrix $\rho_t$ and the matrix $E_t$ and  Eq.(\ref{eq:pqs}) yields the outcome probability distribution expressed in terms of their matrix elements,
\begin{align}
P_p(V) \propto & \rho_{00} E_{00} e^{(-(V-1)^2/2a^2)} + \rho_{11} E_{11} e^{(-(V+1)^2/2a^2)}\nonumber\\
& \quad \quad + (\rho_{10} E_{01}+ \rho_{01} E_{10}) e^{(-(V^2+1)/2a^2)}.\nonumber
\end{align}
We observe that the information obtained after the measurement of interest plays a formally equally important role as the conditional quantum state represented by $\rho$.

The predicted mean value is $\langle V\rangle_p=\int P_p(V) V dV$, and can be evaluated,
\begin{align}
\langle V \rangle_p = \frac{(\rho_{00} E_{00} -  \rho_{11} E_{11}) }{(\rho_{00} E_{00}+   \rho_{11} E_{11} +\exp (-\frac{1}{8a^2})  (\rho_{10} E_{01}+ \rho_{01} E_{10}) }.
\label{eq:mean}
\end{align}
Here we note that if the measurement is  strong,  $a$ is small, and the coherence contribution is cancelled in the denominator, yet if the  
measurement is weak, a single measurement is dominated by noise and reveals only little information (and causes infinitesimal back action). This is the situation that leads to so-called weak values. If the measurement signal is proportional to an observable $\hat{A}$, and the system is initialized in $|\psi_i\rangle$ and post-selected in state $|\psi_f\rangle$, the mean signal is given by \cite{ahar88},
\begin{equation} \label{eq:weakvalue}
\langle \hat{A}_w\rangle = Re[\frac{\langle \psi_f|\hat{A}|\psi_i\rangle}{\langle \psi_f|\psi_i\rangle}],
\end{equation}
which may differ dramatically from the usual expectation value $\langle \psi_i|\hat{A}|\psi_i\rangle$. Our Eq.(\ref{eq:pqs}) has, indeed, been derived by Wiseman \cite{wise02} to clarify how weak values are related to continuous quantum trajectories and correlations in field measurements.


 \begin{figure}\begin{center}
\includegraphics[angle = 0, width =.47\textwidth]{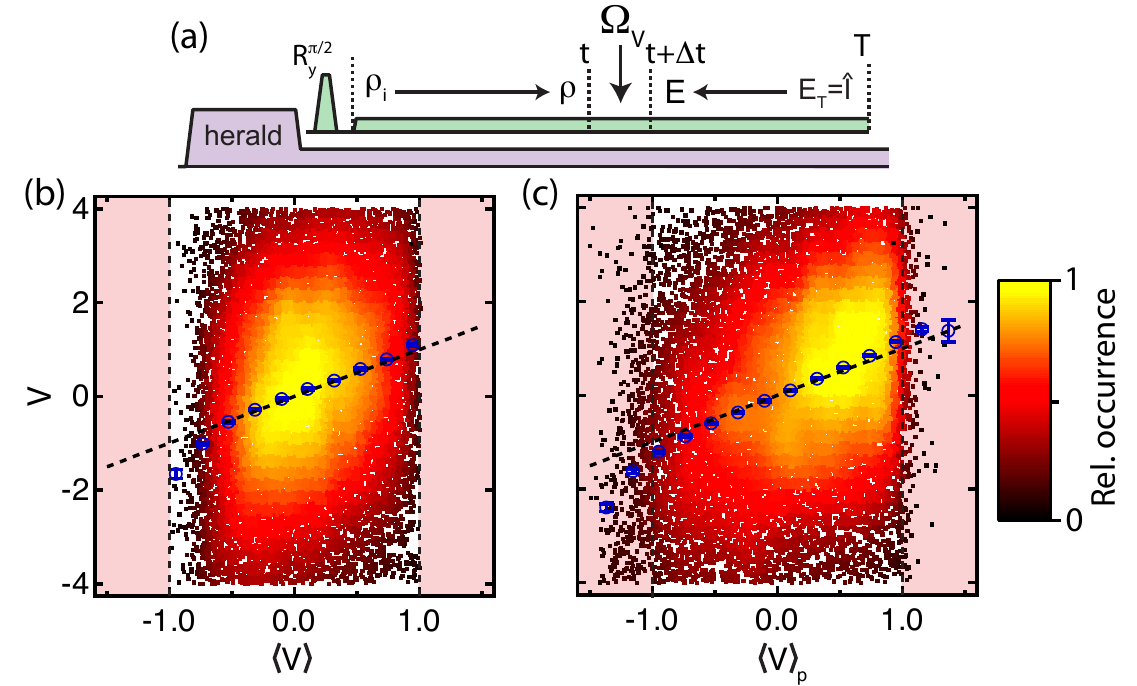}
\end{center}
\caption{ \label{fig4}  Conventional and past quantum state predictions for the measurement $\Omega_V$ conducted at time $t$.  (a) The experiment sequence initializes the qubit along $+x$ and probes the cavity while the qubit transition is driven with a constant Rabi frequency. Each experiment yields a  value $V$ resulting from the  $\Omega_V$ measurement and predicted mean values $\langle V \rangle$ (which is based on $\rho$),  and $\langle V \rangle _p$ (which is based on $\rho$ and $E$).  We plot $V$ versus $\langle V \rangle$ (b),  and $\langle V \rangle _p$ (c) and find that the conditional average of $V$ (open circles) is in agreement with the expected mean value given by the dashed line.  
Note that $\langle V \rangle _p$ makes predictions for the mean value that fall outside of the spectral range of the qubit observable (in the pink region).   \label{vbar}}
\end{figure}

 \begin{figure}\begin{center}
\includegraphics[angle = 0, width =.47\textwidth]{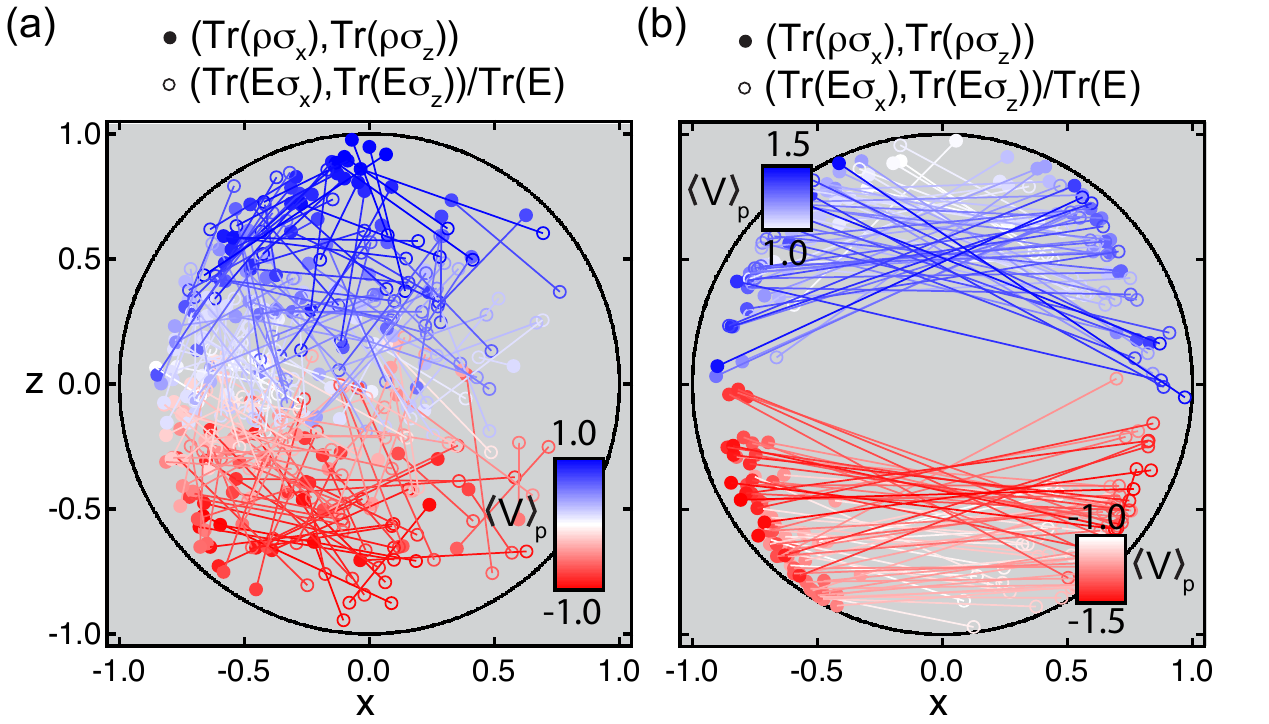}
\end{center}
\caption{ \label{states}   Bloch vector representation of the matrix elements of $\rho$ and $E$.  For each iteration of the experiment, a line joins the coordinates $\{\mathrm{Tr}(\rho \sigma_x), \mathrm{Tr}(\rho \sigma_z)\} = \{\langle \sigma_x\rangle, \langle \sigma_z \rangle\}$ (closed circles) and  $\frac{1}{\mathrm{Tr}(E)}\{\mathrm{Tr}(E \sigma_x), \mathrm{Tr}(E \sigma_z)\}$ (open circles).   The closed circles represents the state of the system at time $t$ based on $\rho_t$, and the open circles represent the corresponding quantity based on $E_{t+\Delta t}$.  The color indicates the value of $\langle V\rangle_p$ for each pair of states.  Panel (a) displays some of the the matrix elements that yield normal predictions ($|\langle V\rangle_p|\leq1$), and panel (b) displays a sample of matrix elements that yield anomalous ($|\langle V\rangle_p|>1$) predictions.} 
\end{figure}

In figure \ref{vbar}, we display results of our experiments that test the predictions of Eq.(\ref{eq:mean}).  For many iterations of the experiment we choose a measurement  time interval, $\Delta t = 180$ ns that is short enough that the effect of the continuous Rabi drive is nearly negligible in the time interval $(t,t+\Delta t)$. Based on $800$ ns of probing before $t$, we calculate $P(V)$, and based on 800 ns of probing before \emph{and} after the measurement interval, we calculate $P_p(V)$ for the result of the measurement. In Fig.\ \ref{vbar}, we show that both the conventional and the past quantum state formalism yield agreement between the predicted mean value and the measured values. The measured results are noisy, and we plot the data with the predicted average value along the horizontal axes, and the measured values along the vertical axes. 

While $\langle V \rangle = \langle \sigma_z \rangle$, and thus  never exceeds $1$, a fraction of the experiments lead to prediction and observation of  values $|\langle V \rangle_p| > 1$.  Such anomalous weak walues in connection with Eq.(\ref{eq:weakvalue}) have been typically identified with the intentional post selection of final states with a very small overlap with the initial state.  Surprisingly, continuous probing leads to similar effects \cite{webe14}. In figure \ref{states} we examine the states that lead to different weak value predictions. We represent pairs of $\rho$ and $E$ as connected points on the Bloch sphere. 
Indeed, predictions outside the spectral range of the operator are accompanied by near orthogonality of states associated with the matrices $\rho_t$ and $E_{t+\Delta t}$.  In agreement with the pure state case, large weak values of $\sigma_z$ do not occur when $\rho_t$ or $E_{t+\Delta t}$ are close to the $\sigma_z$ eigenstates, but rather when they are close to opposite $\sigma_x$ eigenstates.

In conclusion, we have demonstrated the use of the quantum trajectory formalism to infer the quantum state of a superconducting qubit conditioned on the outcome of continuous measurement. We have also demonstrated a quantum hindsight effect, where probing of a quantum system modifies and improves the predictions about measurements already performed in the past.  
These advances may be used to improve the state preparation and readout fidelity for quantum systems and increase their potential for use as probes \cite{mank09,mank09pra,mank11,arme09,whea10,ryba14} of time-dependent interactions and parameter estimation.

*murch@physics.wustl.edu

\begin{widetext}

\section{Measurement calibration}

We calibrate the measurement strength by recording measurement signals for the qubit prepared in the $|0\rangle $ and $|1\rangle$ states. Measurement results are Gaussian distributed and we scale the signal such that the distributions are centered at $+1$ and $-1$ for the $|0\rangle$ and $|1\rangle$ states respectively. The variance,  $a^2=1/4k\eta\Delta t$ is related to the measurement strength $k$, the measurement quantum efficiency $\eta$, and the integration time $\Delta t$. 
\begin{eqnarray}
P(V_t||0\rangle)=\frac{1}{\sqrt{2\pi} a}\exp{-\frac{(V_t+1)^2}{2a^2}},\quad
P(V_t||1\rangle)=\frac{1}{\sqrt{2\pi} a}\exp{-\frac{(V_t-1)^2}{2a^2}}.
\end{eqnarray}

The measurement quantum efficiency is limited by losses in the microwave components and added noise from the amplifiers.  The total quantum efficiency, $\eta_\mathrm{tot} = \eta_\mathrm{col}\eta_\mathrm{amp}\eta_\mathrm{env}$ also includes environmental decoherence of the qubit.  Environmental inefficiency  is small for the parameters of our experiment $\eta_\mathrm{env} = (1+\kappa/8\chi^2 \nbar T_2^*)^{-1} =0.95$, yet we include it in our calculation of   $\rho$ and $E$ which is described below.   

The measurement quantum efficiency $\eta = \eta_\mathrm{col}\eta_\mathrm{amp} $ was measured by fitting the distributions $P(V_t||0\rangle)$ and $P(V_t||1\rangle)$ to determine $a^2 = \kappa/(16 \chi^2 \nbar\eta \Delta t)$, using $\nbar =1.3$ and $\Delta t = 700$ ns. The dispersive coupling rate $\chi/2\pi = - 0.43$ MHz was determined by using a Ramsey measurement to measure both the ac Stark shift  $(2 \chi \nbar)$ and the measurement induced dephasing rate ($\Gamma_m = 8 \chi^2 \nbar /\kappa$) for intracavity photon numbers ranging between $\nbar=0$ and $\nbar = 1.3$.

\section{Calculation of the density and effect matrices}
 The density matrix is calculated by propagating the stochastic master equation for our quantum system \cite{wisebook},
 \begin{eqnarray}
\frac{d \rho}{dt} = -\frac{\mathrm{{\bf i}}}{\hbar} \frac{\Omega}{2} [\sigma_y,\rho]  + (k+\frac{\gamma}{2}) (\sigma_z\rho\sigma_z - \rho)+ 2\eta k(\sigma_z\rho+\rho \sigma_z - 2 \mathrm{Tr}(\sigma_z \rho)\rho )V_t. \label{eq:rho}
\end{eqnarray}
This is the same as equation (4) in the main text  with the exception that we include qubit dephasing which is characterized by the  rate $\gamma=1/T^\ast_2$.  We propagate $\rho$ forward from the initial state $\rho_i = x_i \sigma_x/2 + (\hat{I} +z_i\sigma_z)/2$, where $x_i =0 .864,\ z_i=-0.058$ are determined from quantum state tomography at $t=0$. These values differ from the ideal initial state $\rho_i =\sigma_x/2 + \hat{I}/2$ due to a time delay between the herald measurement and the start of data collection. The Rabi frequency  is $\Omega/2\pi = 0.7$ MHz and we use time steps of $\delta t=20$ ns.
The effect matrix $E(t)$ obeys a corresponding equation \cite{gamm13},
\begin{eqnarray}
\frac{d E}{dt} = \frac{\mathrm{{\bf i}}}{\hbar} \frac{\Omega}{2}[\sigma_y,E]  + (k+\frac{\gamma}{2}) (\sigma_z E \sigma_z - E)+ 2\eta k(\sigma_z E+E \sigma_z -  \mathrm{Tr}(\sigma_z E)E )V_{t-dt}. \label{eq:eprop}
\end{eqnarray}

Here we propagate E \emph{backward} in time from the  final state $E_T = \hat{I}$, and since $\mathrm{Tr}(E)  = 2$, the last terms in  (\ref{eq:rho}) and (\ref{eq:eprop}) are different by a factor of two. Since expressions involving $E$ are normalized, this factor is not essential, but we incorporate it to maintain a consistent value of the trace.

Our previous work \cite{webe14} used a Bayesian argument to calculate the quantum trajectory for the density matrix $\rho$.  In this work, we have focused on application of the stochastic master equation to connect with previous theoretical work. We find that the two techniques are equivalent when the continuous measurement is very weak, and the two techniques give very similar results for the parameters used in this experiment. Figure \ref{sme} displays a comparison of the two methods for a single trajectory.

 \begin{figure}\begin{center}
\includegraphics[angle = 0, width =.6\textwidth]{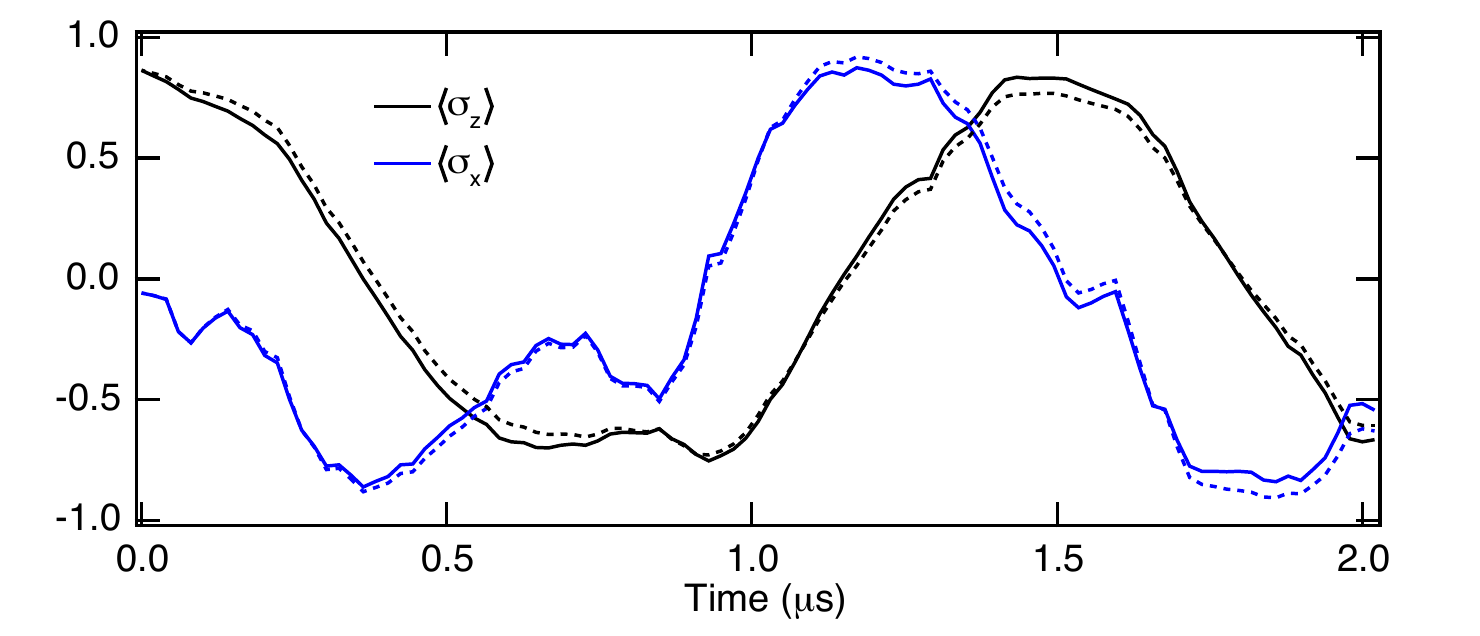}
\end{center}
\caption{ \label{sme} Comparison of the quantum trajectory given by ($\langle \sigma_x \rangle, \langle \sigma_z \rangle$ ) calculated with  Eq.\ref{eq:rho} (solid) and the Bayesian update procedure presented in Ref. \cite{webe14} (dashed).  The two techniques give very similar results. }
\end{figure}

\section{Experimental setup}

Figure \ref{scheme} displays a detailed experimental schematic for the experiment.

 \begin{figure}\begin{center}
\includegraphics[angle = 0, width =.9\textwidth]{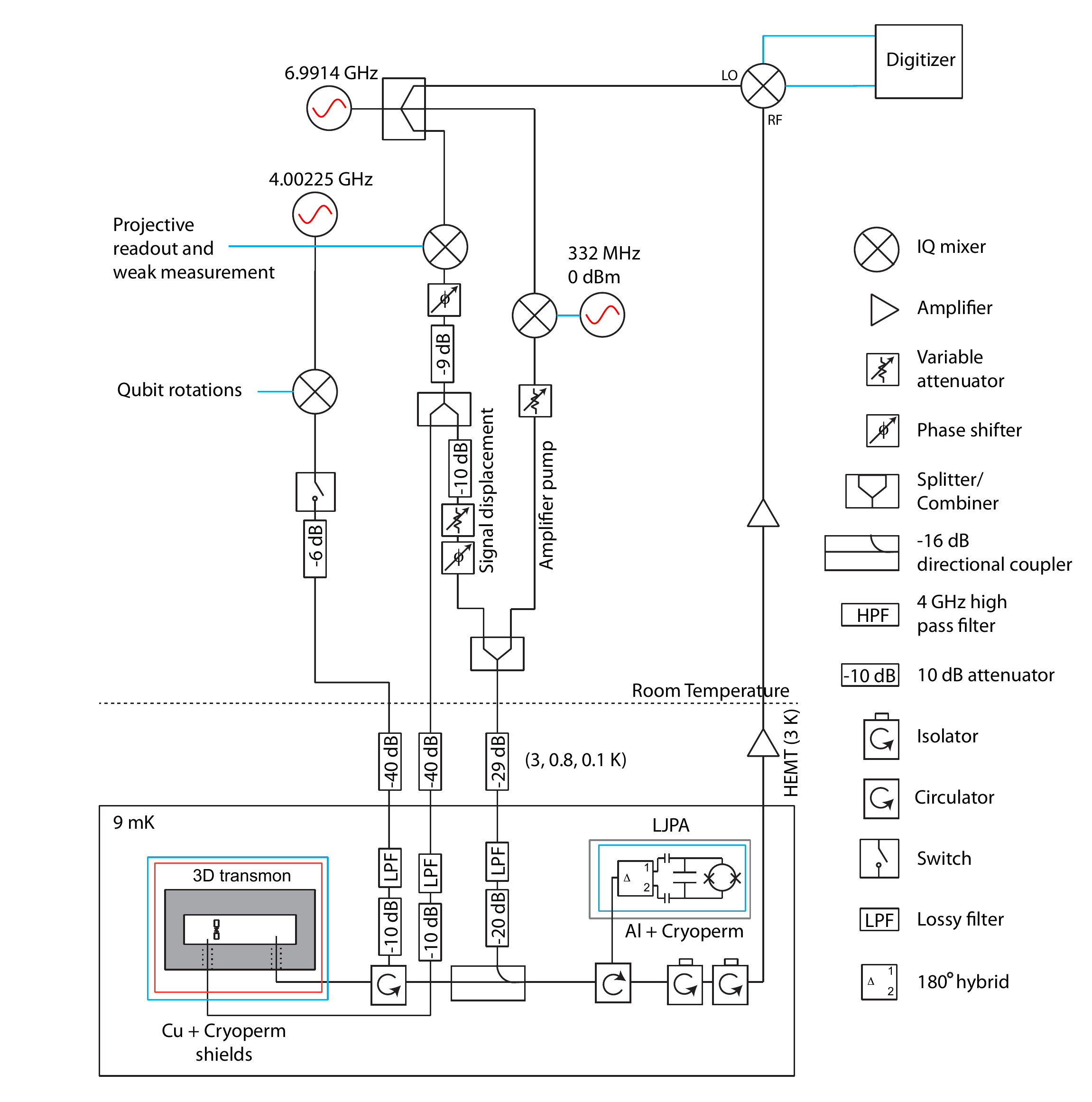}
\end{center}
\caption{ \label{scheme} Experimental schematic.}
\end{figure}

The qubit was fabricated from double-angle-evaporated aluminum on silicon and is characterized by charging energy $E_C/h = 250$ MHz and Josephson energy $E_J/h = 9$ GHz.  We obtain $E_C$ from the anharmonicity of the qubit levels and calculate $E_J$ from the relation $\hbar\omega_q = \sqrt{8 E_J E_C} -E_C$.  The qubit frequency is $\omega_q/2\pi = 4.0033$ GHz, yet all experiments were performed with a constant intracavity photon number $\nbar = 1.3$ which resulted in an ac Stark shift of $1.05$ MHz.  All qubit rotations were performed at the ac Stark shifted frequency, $4.00225$ GHz.  We measured the qubit coherence properties, $T_1 = 30 \ \mu$s, $T_2^* = 16\ \mu$s using standard techniques.

The lumped-element Josephson parametric amplifier (LJPA) is composed of a $1.5$ pF capacitor shunted by a  SQUID composed of two $I_0 = 1\ \mu$A Josephson junctions.  The LJPA is operated with negligible flux threading the SQUID loop and produces 20 dB of gain with an instantaneous bandwidth of $9$ MHz

\section{Conditional averaging} 

The experimental verification of the predictions made by $\rho$ and $E$ in figures 1, 2, and 3 in the main text uses the concept of conditional averaging.  In this section, we describe the procedures used in these figures in greater detail than space allows in the main text. 

In figure 1, we present a trajectory for prediction of the outcome of a projective measurement of the qubit in the $\sigma_z$ basis, based on a single experimental sequence with a duration of $2\ \mu$s between the preparation and the final measurement $M$.  This trajectory is obtained by propagating the stochastic master equation to obtain $\rho(t)$.  We denote this as $\rho_\mathrm{target}(t)$, the target trajectory.   We confirm that the trajectory for $P(+z)$  is correct in the following manner. For each time point $t$ of the trajectory we conduct 3000 experiments with duration $t$ between the preparation and final measurement $M$. Of these 3000 experiments, we examine the subset of experiments that had values $\rho_t$ that were within $\pm 0.02$ of $\rho_\mathrm{target}(t)$. Using this subset, we determine $\tilde{P}(+z)$ based on the corresponding subset of measurements $M$.  

In figure 2, we display a trajectory for the retrodiction $P_p(+z)$ for the measurement $M$. For all of the initial states that we examine, $P_p(+z) = E_{00}/2$.  The trajectory is formed by propagating $E$ for different periods of time (i.e. starting later and later) and we denote this trajectory as $E_\mathrm{target}(t)$.  We confirm that the trajectory $P_p(+z)$ is correct in the following way. For $3\times 10^5$ experimental iterations, we calculate $E(t)$. At each time $t$ we examine the outcomes of measurement $M$ that corresponded to values of $E(t)$ within $\pm0.02$ of $E_\mathrm{target}(t)$ to determine the experimental probability $\tilde{P}(+z)$ conditioned on $E(t)$.

In figure 3, each experimental iteration results in a measurement value $V$ and a predicted average values $\langle V\rangle$, which is based solely on $\rho$ and $\langle V \rangle_p$, which is based on $\rho$ and $E$.  The values $V$ are dominated by noise but we confirm that their average value agrees with the predictions $\langle V\rangle$ and $\langle V \rangle_p$.  We sort the predictions into bins of width 0.2 and average all the measurement values $V$ that have $\langle V \rangle$ or $\langle V \rangle_p$ within each bin. For predictions $\langle V\rangle$ near $-1$ and $\langle V\rangle_p$ near $-1.5$ we note slight deviation from the expected dependence.  Relatively few points contribute to these averages and the deviation may be due to non-Gaussian tails of the distribution.

\section{Retrodiction for different values of $\rho$}

In the main text, figure 2 presents the retrodictions for different initial states $\rho_i$. The initial states $+x$ and $+y$ are prepared by heralding the ground state and applying a resonant rotation to the qubit.  Here we consider two more initial states which are near the eigenstates of $\sigma_z$.  
In figure \ref{heraldz}  we examine the retrodictions that are made for initial states that are prepared in $+z$ and $-z$ which is accomplished by heralding the $+z$ or $-z$ states and applying no rotations. Using the independently measured projective measurement fidelity, we use $\rho_i = \hat{I}/2 + 0.47 \sigma_z$ for the $+z$ preparation and $\rho_i = \hat{I}/2 - 0.46 \sigma_z$ for the $-z$ preparation.  

We note that figure 2 in the main text focuses on the deliberate preparation of the initial state $\rho_i$.  We could have equivalently used conditional dynamics such as the trajectory displayed in figure 1 to prepare different initial states $\rho_i$.  Since previous work has already established conditional preparation of different states $\rho_i$, we chose to focus our attention on the propagation of $E$.

 \begin{figure}\begin{center}
\includegraphics[angle = 0, width =.5\textwidth]{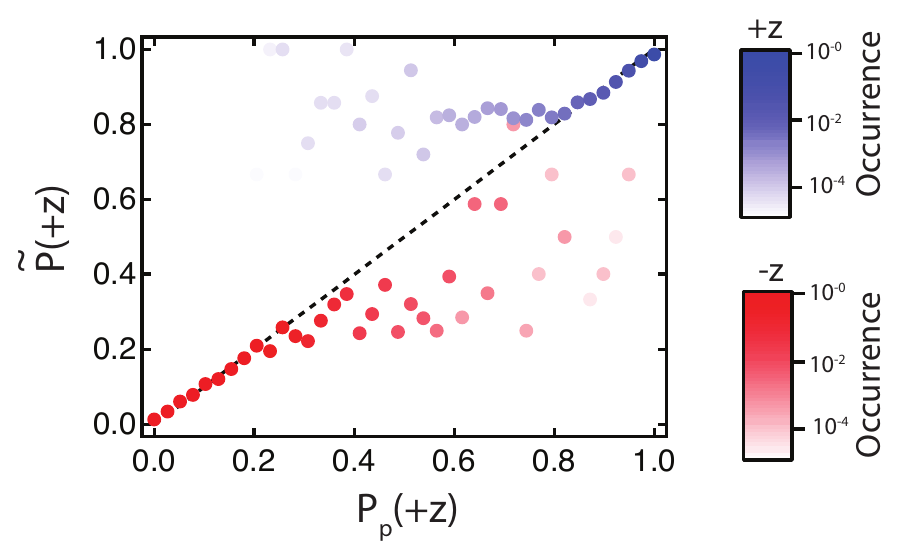}
\end{center}
\caption{ \label{heraldz} Retrodiction for initial states $\rho_i$. Using the same experimental sequence displayed in figure 2 in the main text, with the exception that there is no rotation applied to the state after the herald, we prepare different initial states $\mathrm{Tr}(\rho_i \sigma_z) \simeq \pm 1$. To accomplish this, we use the herald measurement to select either the ground or excited state as the initial state, however the finite measurement fidelity (96\% for $-z$ and 97\% for $+z$ results in imperfect state preparation.  These fidelities are incorporated into $\rho_i$, and thus the retrodiction for measurement $M$.  The majority of the retrodictions give values $P_p(+z)$ near 0 or 1, yet some retrodictions give values between $0$ and $1$ and these retrodictions are confirmed by examining the outcomes of measurements $M$ that yielded similar values of $E_0$.}
\end{figure}

\section{Smoothed predictions of projective measurements}
 \begin{figure}\begin{center}
\includegraphics[angle = 0, width =.6\textwidth]{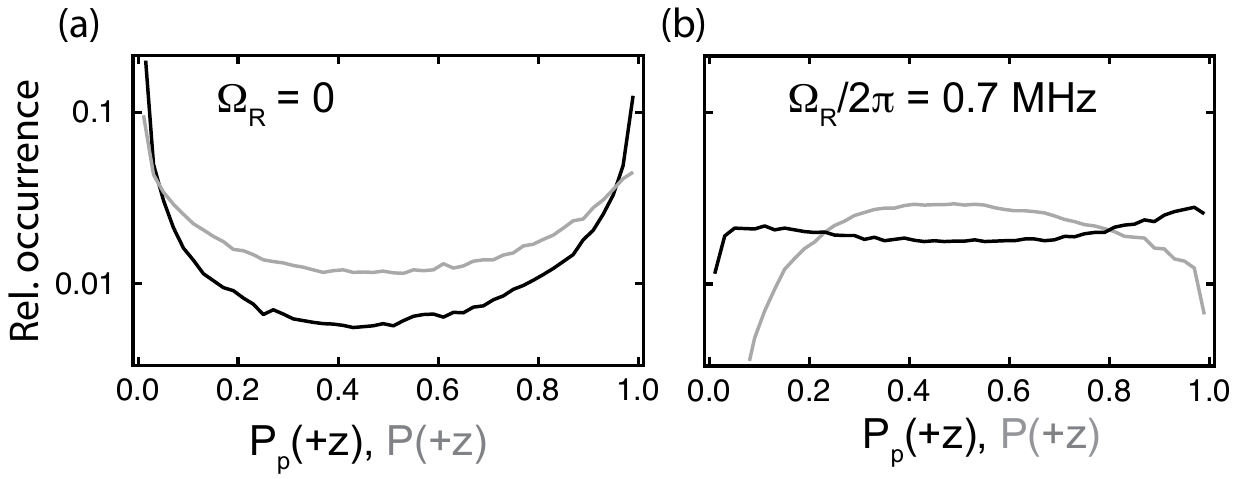}
\end{center}
\caption{ \label{pp0} The occurrence of different values of $P(+z)$ and $P_p(+z)$ obtained from many iterations of the experiment shown in grey and black respectively.  We propagate $\rho$ for 800 ns from the initial state $\mathrm{Tr}(\rho_i\sigma_x) \simeq 1$ and propagate $E$ for 800 ns from the final state $E = \hat{I}$. Panel (a) shows the case where $\Omega_R=0$ and (b) shows the case where $\Omega_R/2\pi = 0.7$ MHz. Both graphs show that the Eq.(2) in the main text makes more confident predictions for the outcomes of projective measurements by more often taking values closer to 0 and 1. }
\end{figure}

By increasing the strength of the measurement $\Omega_V$, we may ultimately perform a projective measurement of the qubit state in the $\sigma_z$ basis.  In figure \ref{pp0}, we display histograms of $P(+z)$ and $P_p(+z)$, indicating the predicted and smoothed probability for finding the qubit in its ground state.  We observe more occurrences of values of $P_p(+z)$ than of $P(+z)$ near $0$ and $1$ indicating that the past quantum state more often makes confident predictions about the outcome of a projective measurement.

When $\Omega_R = 0$ the stronger predictions given by the past quantum state are a consequence of the quantum non-demolition (QND) character of the measurement.  The effects of the measurements commute and the past quantum state analysis merely accumulates measurements in the intervals $(0,t)$ and $(t+\Delta t,T)$.  However, by setting $\Omega_R\neq 0$, we break the QND character of the measurement, and it is necessary to propagate $\rho$ and $E$ using their associated stochastic master equations.


\end{widetext}

\end{document}